# Inflight Performance and Calibrations of the Lyman-alpha Solar Telescope on board the Advanced Space-based Solar Observatory

Bo Chen[1]✉ · Li Feng[2]✉ · Guang Zhang[1]· Hui Li[2]· Lingping He[1]· Kefei Song[1]· Quanfeng Guo[1]· Ying Li[2]· Yu Huang[2]· Jingwei Li[2]· Jie Zhao[2]· Jianchao Xue[2]· Gen Li[2]· Guanglu Shi[2]· Dechao Song[2]· Lei Lu[2]· Beili Ying· Haifeng Wang[1]· Shuang Dai[1]· Xiaodong Wang[1]· Shilei Mao[1]· Peng Wang[1]· Kun Wu[1]· Shuai Ren[1]· Liang Sun[1]· Xianwei Yang[1]· Mingyi Xia[1]· Xiaoxue Zhang[1]· Peng Zhou[1]· Chen Tao[1]· Yang Liu[1]· Sibo Yu[1]· Xinkai Li[1]· Shuting Li[2]· Ping Zhang[2]· Qiao Li[2]· Zhengyuan Tian[2]· Yue Zhou[2]· Jun Tian[2]· Jiahui Shan[2]· Xiaofeng Liu[2]· Zhichen Jing[2]· Weiqun Gan[2]

**Abstract**

The Lyα Solar Telescope (LST) is the first instrument to achieve imaging of the full solar disk and the coronal region in both white light (WL) and ultraviolet (UV) H I Lyα, extending up to 2.5 solar radii (Rs), contributing to solar physics research and space weather forecasting. Since its launch on 9 October 2022, LST has captured various significant solar activity phenomena including flares, filaments, prominences, coronal mass ejections (CMEs). On-orbit observation and test results show that LST covers a continuous spatial range and the wavelengths of 121.6 nm, 360 nm and 700 nm. The Lyα Solar Disk Imager (SDI) has a field of view (FOV) of 38.4′ and a spatial resolution of around 9.5″, while the White-Light Solar Telescope (WST) has an FOV of 38.43′ and a spatial resolution of around 3.0″. The FOV of the Lyα Solar Corona Imager (SCI) reaches 81.1′ and its spatial resolution is 4.3″. The stray-light level in the 700 nm waveband is about $7.8×10^{-6}$ MSB at 1.1 Rs and $7.6×10^{-7}$ MSB at 2.5Rs, and in Lyα waveband it is around $4.3×10^{-3}$ MSB at 1.1 Rs and $4.1×10^{-4}$ MSB at 2.5 Rs (MSB: mean solar brightness). This article will detail the results from on-orbit tests and calibrations.

## 1. Introduction

Space-based solar observations have been conducted for decades. However, imaging observations of the hydrogen Lyman α (Lyα) spectral line, which is crucial for studying the hydrogen-dominated solar transition zone, remain rare. This rarity is largely due to the complex radiation mechanisms in this wavelength, compounded by extensive absorption and scattering effects that obscure changes in radiation brightness and distributions. Additionally, significant challenges in developing detectors, optical components, and radiation calibration arise due to severe absorption and technological complexities in this waveband. As a result, there are few imaging instruments specifically designed for this band, leading to a scarcity of observational data.

Over the past few decades, numerous solar missions have provided high-quality imaging and spectroscopic data. Space-based coronagraphs, dating back to the 1970s, began with the Orbiting Solar Observatory 7 (OSO-7) coronagraph, a pioneer in discovering coronal mass ejections (CMEs). Following OSO-7, several other coronagraphs were deployed before the Solar and Heliospheric Observatory (SOHO). The first coronagraphic observations in the ultraviolet (UV) including the Lyα line were performed by the UV Coronal Spectrometer (UVCS) on board of Spartan 201 (Kohl et al. 1994) who was a prototype of UV Coronal Spectrometer (UVCS) on board the Solar and Heliospheric Observatory (SOHO: Kohl et al., 1995).The Large Angle

Spectroscopic Coronagraph（LASCO）C1, C2, and C3 coronagraphs on board SOHO (Brueckner et al., 1995) have significantly contributed to CME and solar wind research. In 2020, the Solar Orbiter (SolO) was launched, featuring the Metis coronagraph (Antonucci et al., 2020), capable of observing the solar corona in both visible light and hydrogen Lyα wavelengths, with a field of view (FOV) ranging from 1.6° to 3.4°. Few instruments have been developed for imaging the solar disk in the ultraviolet (UV) spectrum, especially in Lyα. The first solar disk Lyα observations date back to the Skylab Apollo Telescope Mount (ATM: Reeves, 1976). With rocket experiments in the 1990s, the Multi-Spectral Solar Telescope Array (MSSTA, Hoover et al., 1991) and the Multiple XUV Imager (MXUVI: Auchere et al., 1999) captured full disk Lyα images. Subsequently, the Very High Angular Resolution Ultraviolet Telescope (VAULT: Korendyke et al., 2001), the Transition Region and Coronal Explorer (TRACE: Handy et al., 1999), and the Extreme-Ultraviolet Imager (EUI: Rochus et al., 2020) on SolO conducted high spatial resolution imaging of Lyα in specific solar regions.

Since 2009, Changchun Institute of Optics, Fine Mechanics and Physics (CIOMP) has conducted extensive research on space X-ray and EUV instruments, which have been applied to the Change'3 (CE-3) EUV camera (Chen et al., 2014), Fengyun-3 D satellite (FY-3 D) Wide-field Auroral Imager (WAI: Zhang et al., 2019), Fenyun-3 E satellite (FY-3 E) Solar X-EUV Imager (Chen et al., 2022) . Based on this foundation, several key technologies have been developed, including stray-light suppression, detection technology, and Lyα radiometric calibration. The LST is capable of stably observing the full solar disk and corona in the Lyα band. Additional details are available in previous articles (Chen et al., 2019). LST has successfully captured numerous full solar disk images in the Lyα and 360 nm wavebands, as well as dual-waveband coronal images in the Lyα and 700 nm bands. Data from the Solar Disk Imager (SDI) and White-Light Solar Telescope (WST) are now accessible at http://aso-s.pmo.ac.cn/sodc/asos.jsp. Based on LST's observing capabilities, its scientific objective and data products have been documented in previous papers (Li et al., 2019; Feng et al., 2019).

## 2. Instrument Overview

LST, an innovative solar telescope in orbit, spans observations from the solar disk to 2.5 Rs and is equipped with four solar imaging channels. It comprises two coronal imaging channels within the Solar Corona Imager (SCI), named SCIUV and SCIWL, which are capable of imaging and observing the corona in the Lyα and 700 nm wavebands, respectively. Additionally, LST features two full solar disk imaging channels that are SDI for the Lyα waveband and WST for the 360 nm waveband. To optimize space utilization and minimize weight, SCI, SDI, and WST are mounted on a common optical bench (Chen et al., 2019). The two SCI channels are positioned on the top of the bench, while the SDI and WST are situated at the bottom, as depicted in Figure 1. Comprehensive specifications of LST are detailed in previous publications (Gan et al., 2023; Chen et al., 2019).

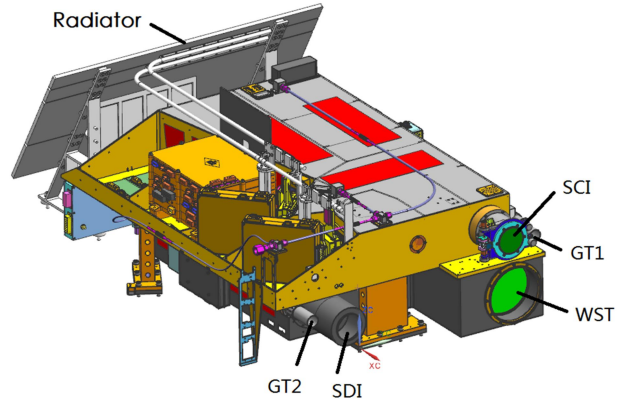

**Figure1** Schematic diagram of LST structure

SCI features a unique design that combines far ultraviolet and visible light in a dual waveband, utilizing an internally occulted reflecting imaging scheme. The SCIUV and SCIWL channels employ the same three-mirror off-axis optical system. A beam splitter in the system reflects the radiation at 121.6 nm while transmitting the visible light at 700 nm, enabling simultaneous imaging of the Lyα and visible light corona. This optical layout is detailed in a previous publication (Chen et al., 2019). The Lyα beam splitter in SCI serves dual functions. First, it reflects Lyα radiation while simultaneously transmitting visible light. Second, it acts as a Lyα reflective filter, characterized by a narrow bandpass and high out-of-band suppression capability. The reflectance of the splitter at the Lyα wavelength is 67.0%, and it has an in-band to out-of-band reflectance suppression ratio of 11.6:1. Its transmittance at the 700 nm visible light wavelength is 93.1%. Constructed from six periodic layers of LaF3/MgF2 multilayer on a fused silica substrate, the splitter features LaF3 and MgF2 layers with thicknesses of 18.0 nm and 13.5 nm, respectively. LaF3 forms the top layer film, while MgF2 is in direct contact with the substrate, forming the bottom layer film.

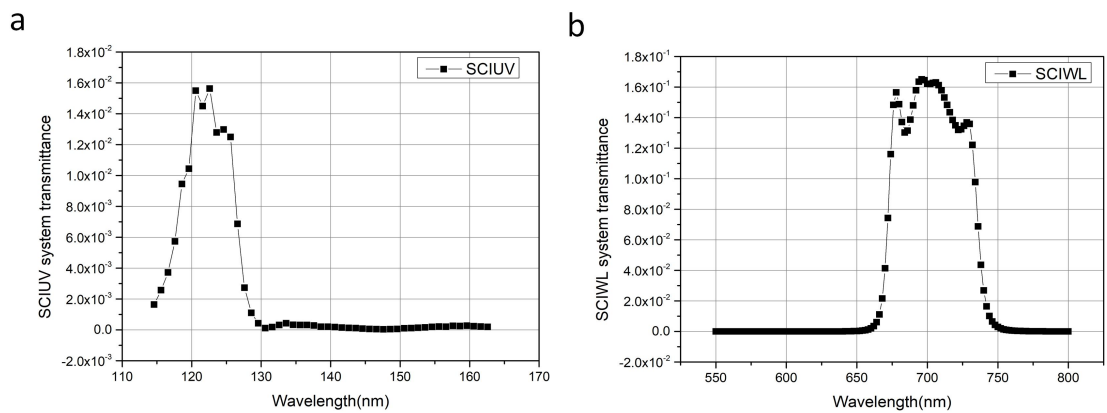

**Figure 2** Spectral response distribution of SCI in 121.6 nm (**a**) and 700 nm (**b**)

Besides the Lyα beam splitter, SCI employs a Lyα filter, which, in conjunction with a narrowband transmission filter and mirror reflectivity distributions, achieves suppression ratios of 93:1 from 115.5nm to 900 nm for 121.6nm and 120:1 from 550 nm to 850 nm for 700 nm. The optical system of SCI exhibits transmittance values of $1.4 \times 10^{-2}$ at 121.6 nm and $1.6 \times 10^{-1}$ at 700 nm, as illustrated in Figure 2.

Both SDI and WST adopt an off-axis two-mirror reflective optical layout, focusing the solar disk on the detectors with an FOV of up to 38′. SDI is dedicated to work in the Lyα line, while WST captures images in the 360 nm waveband. To enhance out-of-band suppression, SDI incorporates three Lyα filters from Acton. Two of these filters are positioned at the entrance of SDI to block out-of-band solar radiation, and the third narrow-band filter is located in the front of the detector. This configuration effectively reduces out-of-band radiation, allowing SDI to capture Lyα radiation with a high spectral purity.

Figure 3a presents the system transmission of SDI. It is characterized by a spectral bandwidth of 9.0 nm and an out-of-band suppression ratio of 50:1. The dual entrance filters of SDI play a crucial role in blocking over 90% of the Sun's intense radiation, significantly contributing to the thermal management of the instrument. This effective thermal control is essential for maintaining the stability and performance of SDI during solar observations.

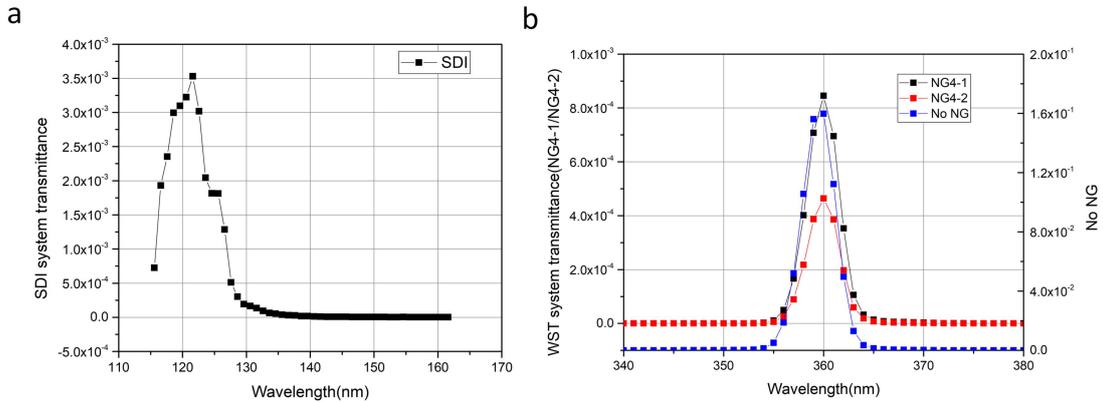

**Figure 3** The system transmission of SDI (**a**) and WST (**b**) for different attenuator windows such as NG4-1, NG4-2 (left coordinate axis) and No NG (right coordinate axis).

WST, operating at the wavelength of 360 nm, is designed to match SDI in both FOV and spatial resolution. At its entrance, WST employs a filter that blocks over 80% of out-of-band radiation, crucial for reducing hot radiation and selectively filtering specific wavelengths. Positioned in front of the CMOS detector, a filter wheel in WST comprises one empty hole and two attenuator windows, offering attenuation levels of $5.3 \times 10^{-3}$ and $2.9 \times 10^{-3}$. These are designed to cover the range of solar irradiation encountered in the 360 nm waveband. The transmittance of the system at 360 nm is $1.6 \times 10^{-1}$, $8.5 \times 10^{-4}$, $4.6 \times 10^{-4}$ for No NG, NG4-1 and NG4-2, respectively, with its spectral response distribution illustrated in Figure 3b.

### 3. On-Orbit Tests

Since the launch of LST, it has captured a substantial number of solar disk and coronal images. These observations have provided insights into various performances of LST. Additionally, specific on-orbit observation modes have been designed to verify other aspects of the performance of the telescope. The on-orbit performance test results have been derived by analyzing the observation and test data. The detailed work on these on-orbit tests and the analysis of the observation results are outlined as follows.

### 3.1 Observations and On-Orbit Tests for SCI

After the self-cleaning of SCI on orbit, it has successfully captured some CMEs and numerous Lyα prominences. Figure 4 showcases images of a prominence and an associated CME. The left panel of the figure displays the prominence in white light, while the right panel shows it in the

Lyα waveband. These images have been obtained by subtracting the respective daily minimum images as background. In the white-light image, features such as the prominence, cavity, and the CME front are distinctly visible. Conversely, in the Lyα waveband, only the prominence is evident. However, by integrating data from both SDI and SCIUV, it has become possible to trace the evolution of the prominence from the solar disk to the coronal region. Ongoing calibration and data processing efforts are expected to refine these observations further. On-orbit observation results indicate that SCI can achieve dual waveband imaging. In addition, a large prominence was observed simultaneously in Lyα and white-light band on 28 July 2023 by SCI. See the supplementary movie for details. In the white-light prominence movie, the bright ring of the original image is removed using a machine learning process, while the Lyα prominence movie is obtained by removing the quiet solar image and a machine learning process to perceptively show the prominence more clearly. The machine learning process is as follows: learn the bright ring information with an on-orbit image sequence, extract the pattern of bright ring appearance, and then remove the bright ring in the new image with the learned pattern.

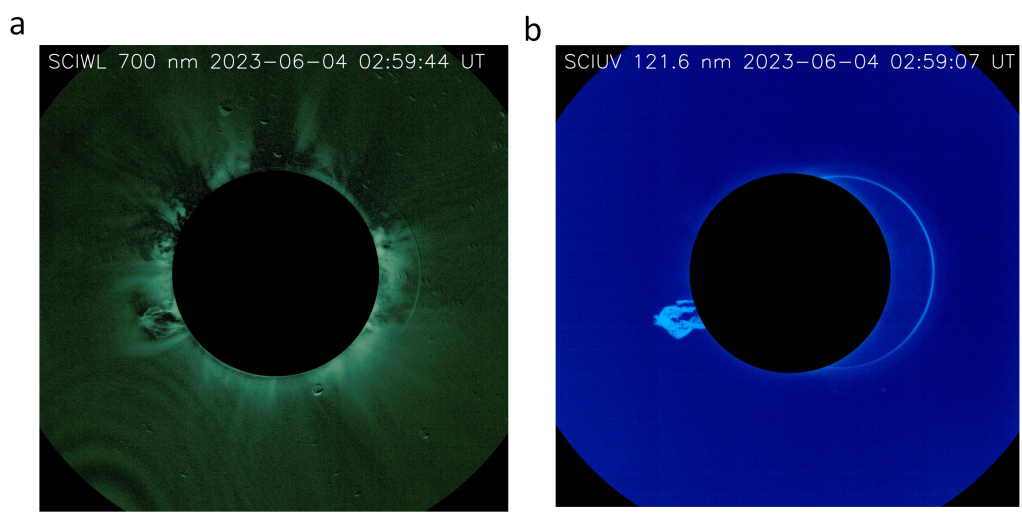

**Figure 4** Images captured by SCI in the dual wavebands of 700 nm (**a**) and 121.6 nm (**b**) simultaneously on 4 June 2023.

In Figure 4 and the movies, the black disks represent the occulted sunlight within 1.1 Rs. The images have been processed by subtracting the daily-minimum background. For the SCIWL channel at 700 nm, the final image is a summation of three images taken at three different polarizer positions. In the polarization wheel of SCI, we installed three polarizers orientated at -60 degree, 0 degree, and 60 degree, respectively (Feng et al., 2019). Subsequently, three such images are summed to create Figure 4a and provide the total brightness of the corona. A different combination of the images would provide also the polarized brightness. The bright ring in the figure results from the illumination of the inner edge of the secondary mirror by the solar edge, and a ghost image is generated after passing through the beam splitter. The position of the bright ring is fixed relative to the center of the secondary mirror. To assess the spatial resolution of SCI, the images of prominences with fine structures captured on 12 June 2023, were selected for analysis, as shown in Figure 5. By examining the minimum prominence structures and measuring the brightness distribution along their radial direction in Figure 5b, the resolution was evaluated based on the full width at half maximum (FWHM) of these structures. The FWHM of the Lyα prominence structure was determined to be approximately 1.8 pixels from Figure 5c. Given the pixel size of 2.4″, this equals to a spatial resolution of 4.3″ and SCIWL has a similar result. Thus, the spatial resolution of SCI is determined to be better than its designated resolution of 4.8″.

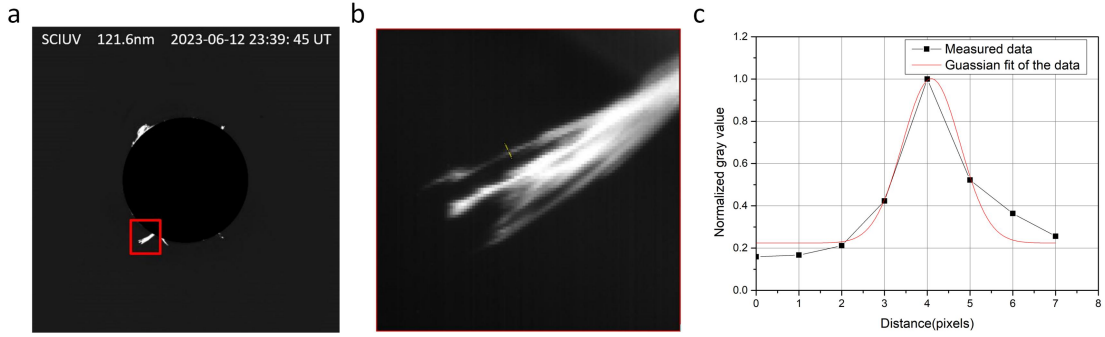

**Figure 5** Lyα prominence image observed by SCIUV on 12 June 2023 (**a**), enlarged image inside the red line (**b**), and normalized gray distribution of the fine prominence structure (**c**).

To evaluate the stray-light suppression level of SCI, the coronal distribution was measured and compared with the solar disk brightness at the same time, leading to the determination of the stray-light suppression distribution of SCI. For the SCIWL channel, the distribution of stray-light suppression level was ascertained using both the visible corona image and the mean brightness of the solar disk (mean solar brightness, MSB), as depicted in Figure 6. The results reveal that the average stray-light level at 1.1 Rs and 2.5 Rs are approximately 6.1 to $7.8\times10^{-6}$ MSB and 6.2 to $7.6\times10^{-7}$ MSB, respectively, with two different approaches. However, it was observed that the stray-light suppression capability of SCI is worse than the on-ground measurement results in previous articles (Gan et al., 2023). This can be primarily attributed to the illumination of the inner edge of the secondary mirror by the solar edge, resulting in a bright ring at the inner edge of the FOV. This phenomenon adversely affects the stray-light suppression performance of SCI. Refer to the Appendix A for the detailed calculation procedure.

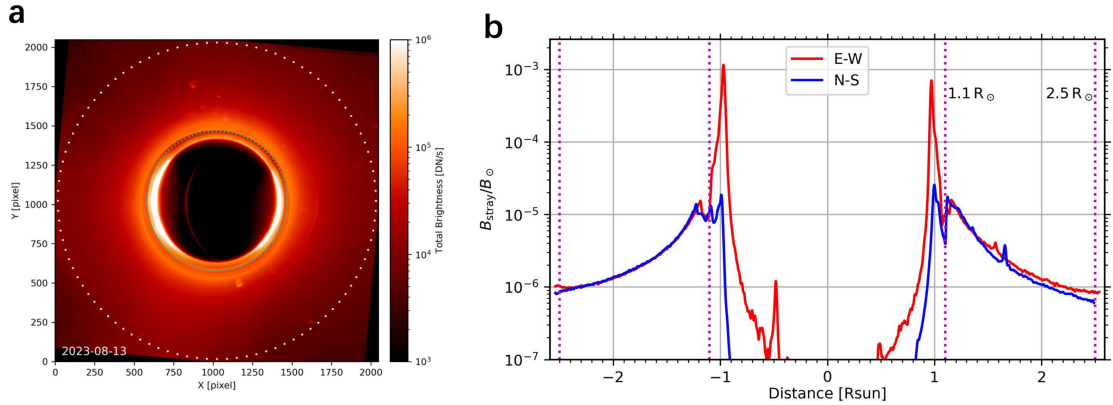

**Figure 6** (**a**) The minimum image of the eight measurements rotated every 45 degrees with respect to the Sun-satellite line by SCIWL. (**b**) SCIWL stray-light distribution along the east-west and north-south directions crossing the FOV center.

In addition, the stray-light level of SCIUV is evaluated by the Lyα corona image captured on 8 July 2023. It was calculated using the captured Lyα image combined with the average brightness data of the solar disk, as depicted in Figure 7. In this analysis, the solar disk brightness in the Lyα band is referenced from SDI at the same time. The resulting stray-light distribution along the yellow line is presented in Figure 7. The average stray-light measured at 1.1 Rs and 2.5 Rs are approximately $5.5\times10^{-3}$ MSB and $5.3\times10^{-4}$ MSB, respectively. The standard deviations are

$3.7×10^{-3}$ MSB at 1.1 Rs and $3.1×10^{-4}$ MSB at 2.5 Rs. Refer to the Appendix B for the detailed calculation procedure.

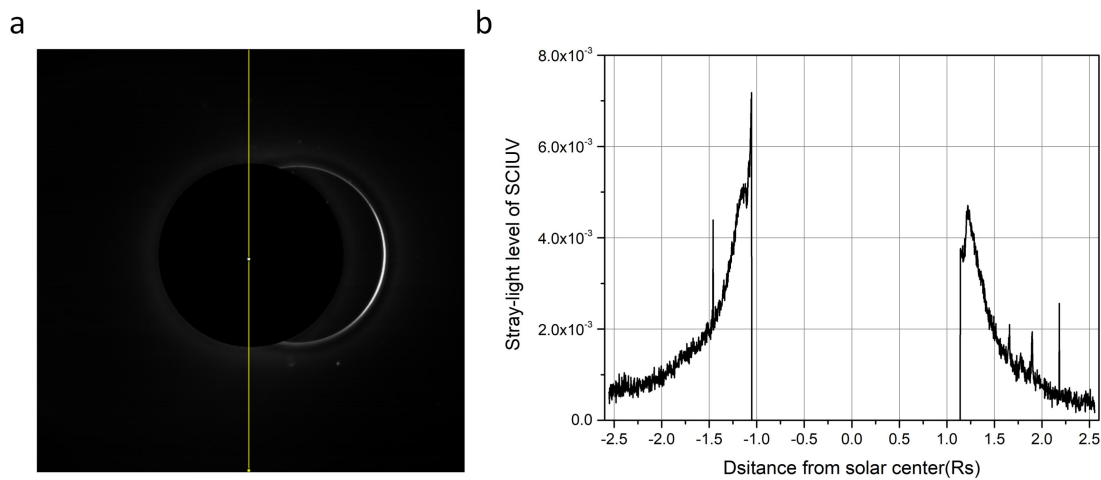

**Figure 7** Image of SCI WL channel (**a**) and stray-light distribution along the yellow line (**b**)

### 3.2 Observations and On-Orbit Tests for SDI

SDI has captured a number of solar active phenomena in the H I Lyα line, including flares and eruptive prominences. In Figure 8, an eruptive prominence observed on the south-east limb on 4 June 2023 is shown. The image has undergone several processing steps, including dark current and flat field calibration, rotation to align with solar north, and a de-spiking procedure to remove hot pixels and cosmic ray interference. Additionally, the image reveals bright active regions and dark filaments across the solar disk. It is observed that filaments in the Lyα waveband appear more diffuse than those typically seen in the Hα waveband. SDI has the capability to shift between routine mode and burst mode. When a local enhancement is detected, for instance, a flare, SDI changes its cadence from 1 minute to 6 seconds, and data in a cutout window of 1024 × 1024 pixels are transmitted to the ground. More details of the triggering and termination scheme for LST instruments can be found in Lu et al. (2020).

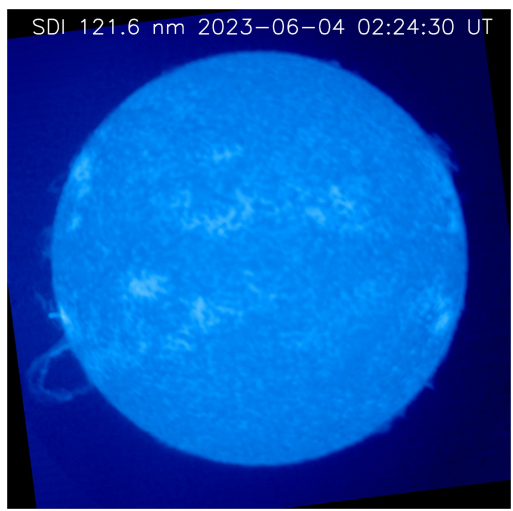

**Figure 8** A Lyα full-disk image with an eruptive prominence at the south-east limb observed by SDI on 4 June 2023.

The results from on-orbit observations indicate that the FOV of SDI is consistent with the design specification. The imaging range of a solar radius in SDI is calibrated by the size of the

Sun on that day, and then the FOV is calculated according to the effective pixel diameter of SDI imaging. The FOV of SDI is calculated by the Sun size on 25 June 2023.The corresponding effective FOV is 38.4'.

However, the observed spatial resolution of SDI appears to be lower than designed. Owing to the lack of sufficiently fine features in the Sun's Lyα waveband, the fast Fourier transform (FFT) method has been employed for evaluating the resolution of SDI. According to this evaluation, the spatial resolution is approximately 9.5" for the first-light images on 26 October 2022, with more detailed information available in Appendix C. The primary cause of this deviation may be identified as the wavefront errors of the entrance Lyα filter, leading to defocusing and image blurring.

### 3.3 Observations and On-Orbit Tests for WST

WST has acquired a substantial number of images at 360 nm and successfully captured more than 50 white-light flares since the launch of the ASO-S satellite (Jing et al., 2024). These flares are distinguished by an enhancement in the Balmer continuum of the hydrogen atom. The results from on-orbit observations indicate that the FOV of WST is also consistent with the design specification. The WST FOV test method is the same as that of SDI and is calculated by the Sun size on 16 March 2023. The corresponding effective FOV for WST is 38.4'.

Figure 9 shows a calibrated full-disk image captured by WST, featuring a white-light flare indicated by intensity brightening, marked by the arrow in the enlarged square. For an in-depth analysis of the physical processes of these white-light flares as observed by WST, refer to Jing et al. (2024) in this topical collection. The on-orbit observation results demonstrate that the FOV of WST is consistent with the design specifications, and it is capable of capturing full solar disk images. WST also has a burst mode which obtains the triggering signal from that of SDI. In the burst mode, WST has a time cadence of one second for the first five minutes and two seconds for the second five minutes and has a cutout window of 1024×1024 pixels as well.

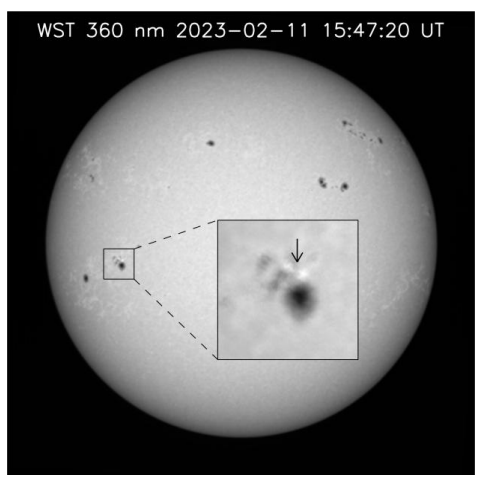

**Figure 9** A 360 nm full-disk image with an enlarged white-light flare observed by WST on 11 February 2023

To assess the resolution, an image of a tiny sunspot light bridge structure captured by WST was analyzed. This structure, observed within a sunspot on 06 March 2023, and shown in Figure 10, was used to determine the spatial resolution. By extracting the smallest bright bridge feature and analyzing the brightness distribution along its vertical direction, the spatial resolution was calculated based on the FWHM. The FWHM of this bright feature is approximately 6 pixels.

Given that each pixel corresponds to a resolution of 0.5", the measured structure corresponds to an angular size of approximately 3.0". The image in Figure 10 illustrates that the spatial resolution of WST is lower than the expected 1.2". A significant factor contributing to this reduced resolution is a temperature gradient of approximately 2 ℃ between the center and edge of the entrance filter when subjected to intense sunlight irradiation. This temperature differential causes a wavefront change within the entrance filter, maybe resulting in defocusing and image blurring.

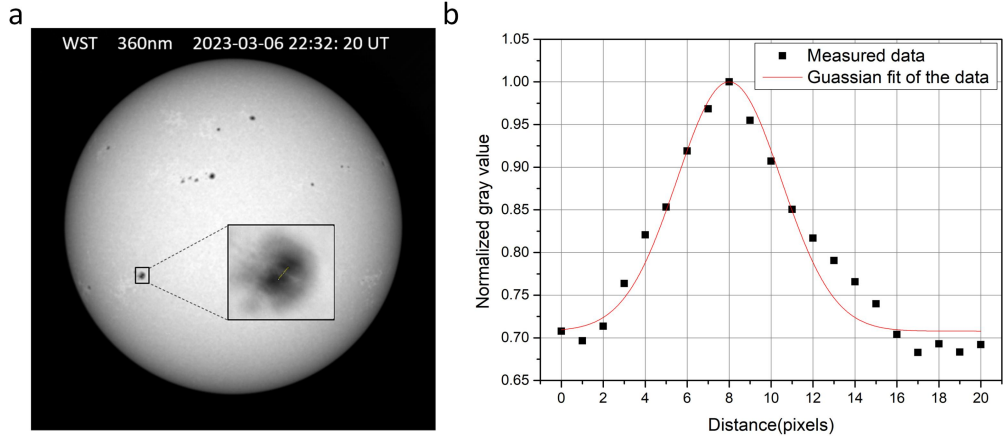

**Figure 10** The sunspot light bridge observed by WST on 06 March 2023 (**a**) and the distribution map of the structure along the yellow line (**b**).

In summary, we have conducted on-orbit tests for over a year and completed most of the planned tests. While some measured performance have met the design requirements, three other performances, including the spatial resolution of SDI, the spatial resolution of WST and the stray-light level of SCIWL（that influences also the performance of SCIUV）are lower than expected and require further improvements and measurements.

## 4. On-Orbit Calibrations

We describe the LST on-orbit calibrations for WST and SDI. Unfortunately, SCI does not work as expected, we are still working on the dark current, flat field, and radiometric calibrations for SCI.

### 4.1 Dark-Field Calibrations of WST and SDI

To derive the dark fields of WST and SDI for their working temperature and exposure time, the dark current models are devised for WST and SDI. Those models are functions of detector temperature and exposure time. For WST, the CMOS detector works at -25±0.2℃. The temperature sequence has seven values and is centered at -25 ℃, i.e. (-30，-28，-26，-25，-24，-22，-20) ℃. The exposure time sequence has 30 values from 0.003 to 1.0 s.

In order to decrease the contamination rate and extend the time interval of the CMOS clearing, the CMOS detector of SDI mostly works at -15±0.2 ℃. The temperature sequence spans (-20, -18, -16, -15, -14, -12, -10) ℃, and the exposure time covers from 0.0019 to 120s with a total of 30 values as well. Figure 11 shows an example of the SDI dark current measurements as a function of detector temperature and exposure time.

Based on these measurements, the dark current models are established. With several tests, we find that the dark field is best fitted with a function such as $I_{dc}=a_0 y+a_1+ (a_2 y^2+a_3 y+a_4) x$, where y is the temperature of the detector, x is the exposure time minus the minimum exposure that is used for obtaining the bias of the detector, and $I_{dc}$ is the measured counts of the dark current. Subsequently the dark field for the detector temperature and exposure time for each image acquisition can be derived. In Figure 12, the dark field data are obtained which are around the working detector temperature and exposure time and the corresponding dark field models. The upper and lower panels show the results for WST and SDI, respectively. We find that the model can recover the observed dark field with median difference values of 5.3 DN and -0.37 DN for WST and SDI, respectively.

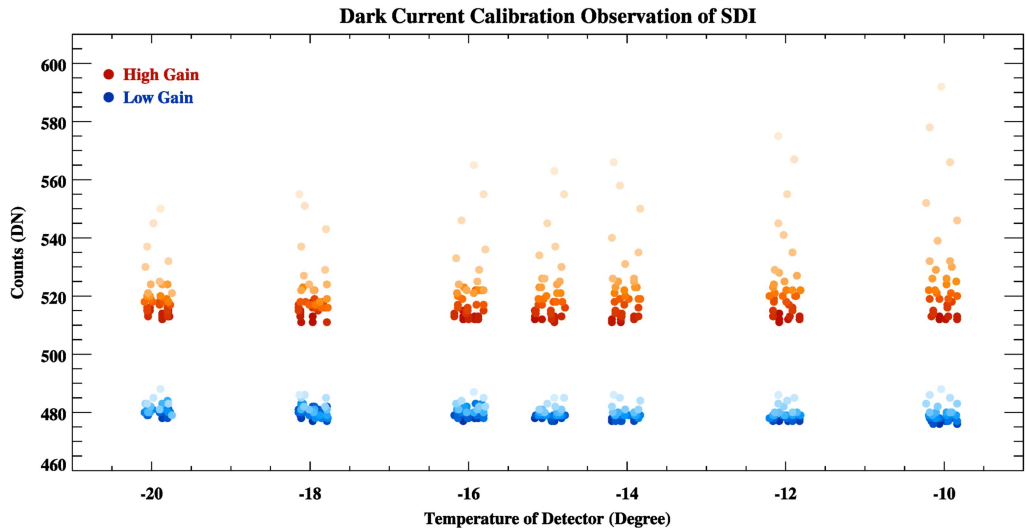

**Figure 11** Measured dark current counts as a function of detector temperature and exposure time for SDI. Red and blue symbols are used for the CMOS high-gain and low-gain modes, respectively. The counts for different exposure times from low to high values are indicated by dots from deep to light colors.

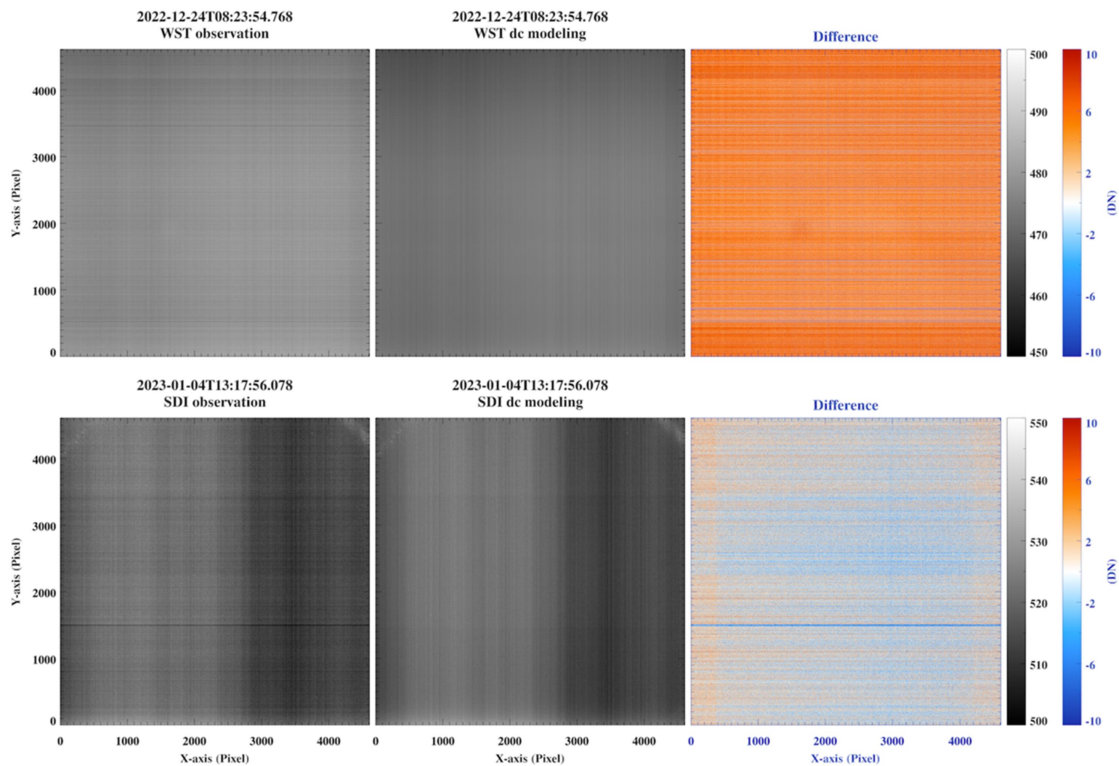

**Figure 12** Comparison of the observed dark field around the working detector temperature and exposure time and the corresponding dark field model for WST (upper panel) and SDI (lower

panel), respectively. The difference is the observed dark field minus the modeled dark field.

**4.2 Flat-Field Calibrations of WST and SDI**

WST and SDI flat fields have been measured in flight by using some different methods. SDI has three flat-field calibration methods which include two KLL (Kuhn, Lin and Loranz, for short KLL, Kuhn et al., 1991) methods and one light-emitting diode (LED) monitoring method. One KLL method off-points the instrument from the Sun-Earth line by the spacecraft with relatively large displacements (for short SAT-KLL method), and the other KLL method with small displacements using the piezoelectric table (PZT, for short PZT-KLL method). The LED-monitoring method uses a LED in front of the filter to illuminate the detector for monitoring its attenuation. WST has two in-flight flat-field calibration methods, one is the normal SAT-KLL method and the other is the LED method (Li et al., 2021).

We adopt the proposed 21 positions by Li et al. (2020a, 2020b) so that the in-flight flat fields of SDI and WST can be done simultaneously. A set of data takes about 80 minutes which is the time required for the attitude adjustment and stabilization of the satellite as well as image acquisition. Three images with the same exposure time were collected from each position, we used all these images for the 21 positions and then used the 21 median images to produce a flat field.

For WST, the Sun is an ideal source for the in-flight flat-field calibration with the SAT-KLL method. The full-disk photosphere images in general show little temporal evolution during the calibration. A sample in-flight KLL flat-field for WST is shown in Figure 13. The WST flat field has many stripe structures due to laser annealing of the CMOS. The precision of the flat field in each pixel is defined as the ratio of the standard deviation to the mean value for a few repetitive measurements. Subsequently, the precision of the flat field in the designed FOV within 1.2 Rs is the average of the precision of all pixels in the FOV. It is approximately 0.5% for WST, which is better than the results from the simulated data in Li et al. (2021).

The SDI in-flight flat field shows more the evolutions of the solar structures than the flat field of WST. A sample in-flight flat field for SDI is shown in Figure 14, the SDI flat field also has many stripe structures due to laser annealing of the CMOS. The red dashed line in the right panel circles the area with reliable flat field measurements and defines the effective FOV of SDI. The in-flight flat field precision is around 2.5%.

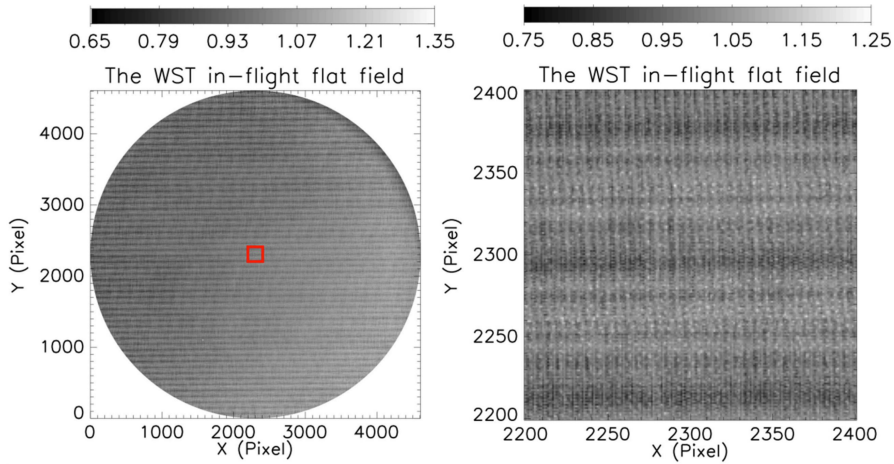

**Figure 13** The in-flight flat field on 4 January 2023, in the nominal FOV of WST (left) and partial image in the red box shown on the left (right).

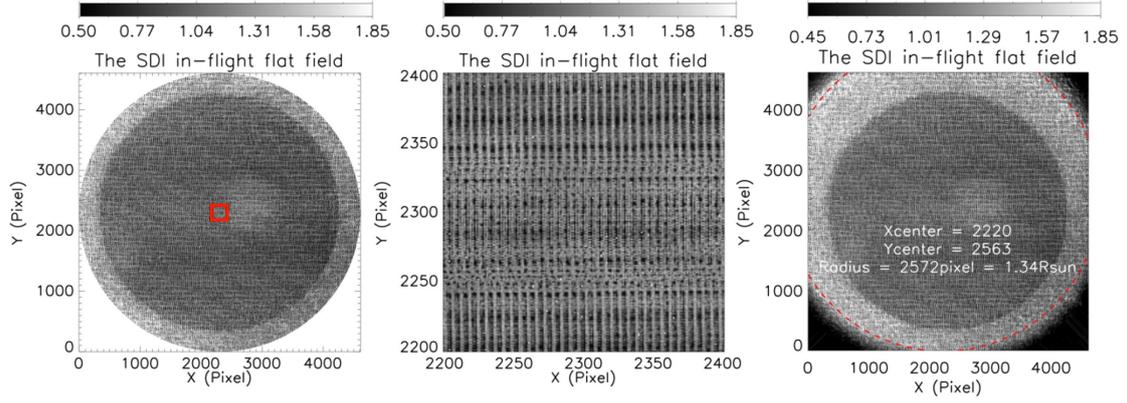

**Figure 14** The high gain in-flight flat field on 4 January 2023 in the nominal FOV of SDI (left) and partial image in the red box shown on the left (middle). The red dashed line in the right panel circles the area with reliable flat field measurements.

### 4.3 Radiometric Calibrations of WST and SDI

Different methods are used to calibrate the observed fluxes of WST at 360 nm and SDI in Lyα. For WST, we use a standard reference solar spectral irradiance at 360 nm from the American Society of Testing and Materials (ASTM G173-03) to calibrate the flux. For SDI, we use the calibrated solar flux in Lyα observed by the SOlar Stellar Irradiance Comparison Experiment (SOLSTICE) on the Solar Radiation and Climate Experiment (SORCE) or by the Extreme Ultraviolet Sensor on board the Geostationary Operational Environmental Satellite (GOES/EUVS) that has a similar waveband as SDI to calibrate the SDI flux.

ASTM G173-03 tables provide a standard reference solar spectral irradiance at 280-4000 nm. Table 1 lists the reference spectra from 358 to 362 nm (360±2 nm, i.e., the waveband of WST), with a constant interval of 0.5 nm. Integrating these irradiances over the wavelength we then get a standard flux of $3.7349\times10^3$ erg cm$^{-2}$ s$^{-1}$ at 360 nm. By using the observed flux of $(3.6043\pm0.0045)\times10^{12}$ (for filter 1 with a transmittance of ≈0.0053) and $(1.8342\pm0.0029)\times10^{12}$ DN s$^{-1}$ (for filter 2 with a transmittance of ≈0.0029) during November 2022, we can obtain the radiometric calibration factors of $(1.0362\pm0.0013)\times10^{-9}$ and $(2.0362\pm0.0032)\times10^{-9}$ erg cm$^{-2}$ DN$^{-1}$ for the two filters of WST, respectively. As shown in the upper panel of Figure 15, the calibration factor increases with time from August 2023 to January 2024, mainly due to the degradation of WST. The precision is defined as the standard deviation of the calibration factors in a few weeks which is about 2%.

**Table 1** ASTM G173-03 reference spectra around 360 nm

| Wavelength (nm) | Irradiance (erg s$^{-1}$ cm$^{-2}$ nm$^{-1}$) |
|---|---|
| 358.0 | 7.8916E+2 |
| 358.5 | 7.3100E+2 |
| 359.0 | 8.5805E+2 |
| 359.5 | 1.0321E+3 |
| 360.0 | 1.0890E+3 |
| 360.5 | 1.0265E+3 |
| 361.0 | 9.4150E+2 |
| 361.5 | 9.1800E+2 |
| 362.0 | 9.5800E+2 |

The radiometric calibration of SDI is done with an indirect method by using the

GOES/EUVS irradiance in the Lyα passband. The lower panel of Figure 15 presents the calibration factor as a function of time. Long-term variation of the degradation of SDI has been monitored and then the radiometric calibration factors are corrected during the lifetime of ASO-S. Note that there is a periodic variation with a period of about 25 days in the time series of the radiometric calibration factors of SDI. This could be explained by two reasons: (1) there exists a periodic variation of solar rotation in the transition region represented by the Lyα irradiance (Zhang et al., 2023); (2) the wavelength ranges plus response functions of SDI and GOES/EUVS are not exactly the same, which could not remove the periodic effect of the solar rotation in the transition region.

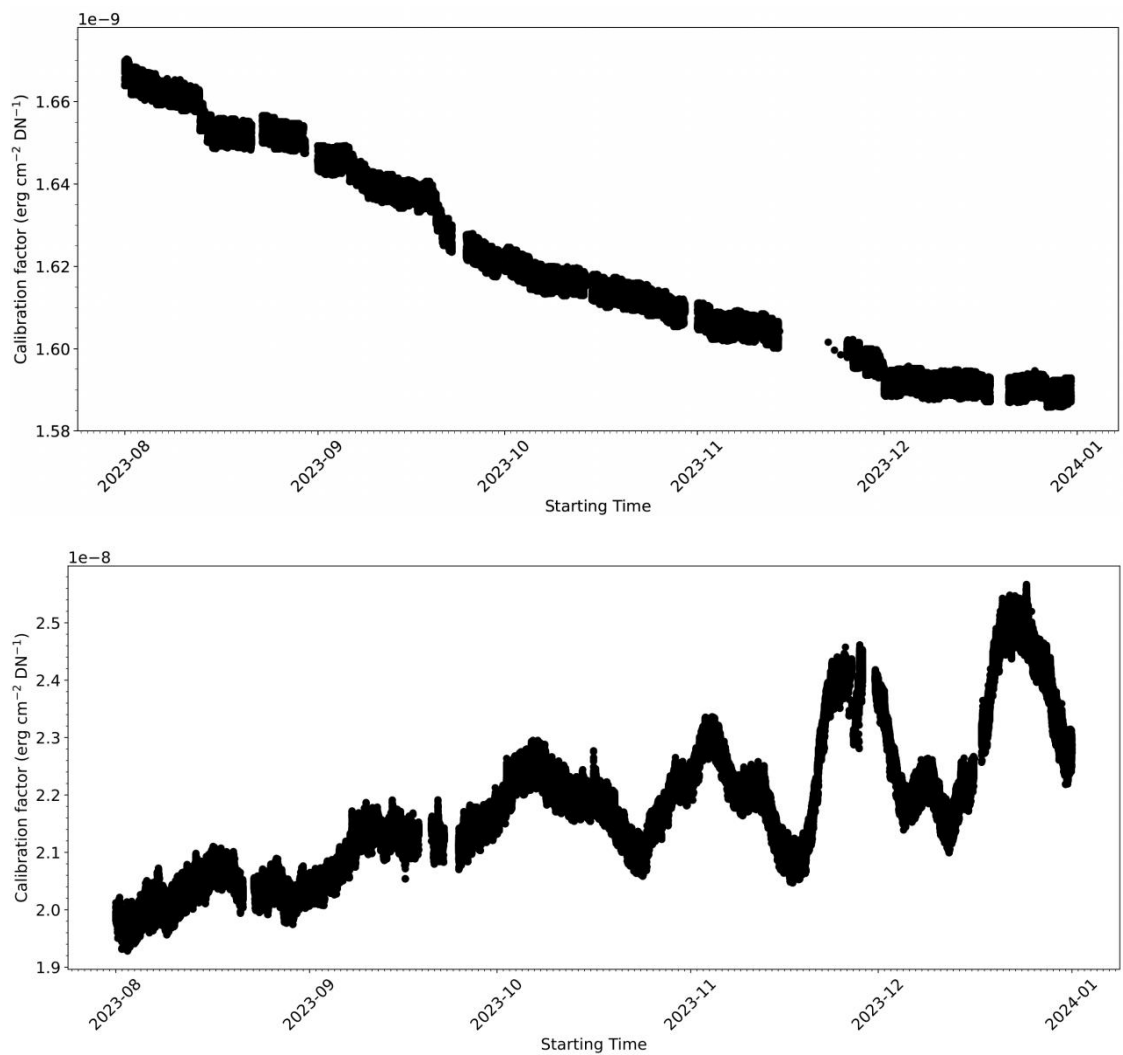

**Figure 15** WST (upper panel) and SDI (lower panel) radiometric calibration factor as a function of time.

## 5. Discussion

LST has been working in orbit for over a year and has successfully undergone tests. However, three specifications have not met expectations yet, necessitating further on-orbit testing and adjustments to enhance image resolution, minimize stray-light influence, and achieve higher quality images.

For SDI and WST, the primary factors affecting resolution could be wavefront changes in the entrance filter, leading to system defocus and the deterioration of the point spread function (PSF). Therefore, the image deconvolution can be used to improve the quality of degraded images. At present, the image quality can be significantly improved by extracting the PSF of SDI from the on-orbit images. To refine the accuracy of deconvolution, the next step involves conducting further on-orbit testing to obtain a high-accuracy PSF for the entire image plane, thus improving the overall image quality. In addition, adjustments to LST's orientation are planned to capture images of UV stars with known brightness for extended exposure radiometric brightness calibration, aiming to enhance the precision of radiometric calibration.

Regarding SCI, additional processing is required to minimize the effects of the bright rings, reducing their impact on coronal imaging. Moreover, SCI and SDI share an overlapping FOV between 1.1 Rs and 1.2 Rs. By comparing images of solar prominences captured in this region by the two instruments, it is theoretically possible to obtain SCIUV images with an absolute brightness. Adjustments in satellite orientation will also be made to image UV stars with known brightness, possibly enabling radiometric brightness calibration for SCI.

## Acknowledgement
The ASO-S mission is supported by the Strategic Priority Research Program on Space Science, the Chinese Academy of Sciences (CAS).

## Author Contributions
Bo Chen and Li Feng contributed equally to this work, are co-first authors and co-corresponding authors listed in no particular order.

LST team of CIOMP carried out the ground spectral response calibrations, some in-flight performance measurements and in-flight calibrations. Bo Chen proposed the relative in-flight measurement and calibration scheme. Some persons proposed in-flight operation models, carried out in-flight performance measurements, calibrations, and the data processing algorithm, including Kefei Song, Lingping He, Guang Zhang, Quanfeng Guo, Haifeng Wang, Shuang Dai, Xiaodong Wang, Shilei Mao, Peng Wang, Kun Wu, Shuai Ren, Liang Sun, Xianwei Yang, Mingyi Xia, Xiaoxue Zhang, Pen Zhou, Shen Tao, Yang Liu, Sibo Yu. Some persons carried out the ground spectral response calibrations, including Lingping He, Haifeng Wang and Xiaodong Wang. Guang Zhang wrote the relevant part of the manuscript.

The LST team at Purple Mountain Observatory (PMO, led by Li Feng and Hui Li) planned and carried out different in-flight calibration modes and corresponding analyses for in-flight dark current (Jie Zhao, Shuting Li), flat field (Jingwei Li), radiometric calibrations (Ying Li, Gen Li), etc. Yu Huang submitted all the in-flight test commands to the ground segment which subsequently were transmitted to the spacecraft. The PMO team also developed the data processing pipe line to produce higher-level scientific data (Li Feng, Ying Li, Lei Lu, Beili Ying and all the others) for the evaluation of the in-flight performance of WST, SDI, and SCI, including SCI stray-light level estimates (Jianchao Xue), spatial resolution (WST: Dechao Song, SDI: Guanglu Shi, SCI: Beili Ying), field of view (WST: Qiao Li, SDI: Beili Ying, SCI: Lei Lu), etc. Weiqun Gan as PI of the ASO-S mission supervised all the activities.

## Funding

This work is supported by the Strategic Priority Research Program of the CAS (Grant No. XDB0560000); National Natural Science Foundation of China (Grant No.12233012, 11921003, 12273040); National Key R&D Program of China (Grant No. 2022YFF0503003, 2022YFF0503004 ,2022YFF0503000); the Joint Research Fund in Astronomy (U2031122) under cooperative agreement between the National Natural Science Foundation of China (NSFC) and Chinese Academy of Science (CAS).

**Declarations**

**Competing interests** The authors declare no competing interests.

**Appendix**

**A. Estimate of the SCIWL Stray-Light Level**

The stray-light suppression level is the ratio of the stray-light brightness to the mean solar brightness (MSB). The stray-light brightness is approximated by the minimum image of the measurements rotated every 45 degrees with respect to the Sun-satellite line. The MSB is calculated by using the solar disk image of SCIWL on 22 March 2023, when part of the solar disk entered the field of view (FOV) of SCIWL, and the two sharp corners in the solar image were not saturated (see Figure A1). The secondary mirror is made of gradient reflective film near the internal FOV, and its reflectivity is about 10% of that of the non-gradient region. By fitting the edge of the solar disk (green dotted line), using the brightness at the plus signs, the limb darkening model at 700 nm, and the reflectivity of the second mirror, it is estimated that the MSB is about $1.1 \times 10^{10}$ DN/s with a standard deviation of $6.9 \times 10^{8}$ DN/s. The stray-light levels of SCIWL can be obtained as the ratio of the brightness of the stray-light image to the MSB. The average stray-light levels at 1.1 Rs and 2.5 Rs are approximately $7.8 \times 10^{-6}$ MSB and $7.6 \times 10^{-7}$ MSB, respectively, with the standard deviation of $5.1 \times 10^{-7}$ MSB at 1.1 Rs and $5.0 \times 10^{-8}$ MSB at 2.5 Rs. Alternatively, we evaluate the stray-light levels with a second method by converting the stray image from DN to MSB using the radiometric calibration on the ground. It turns out that the stray-light levels at 1.1 Rs and 2.5 Rs are about $6.1 \times 10^{-6}$ MSB and $6.2 \times 10^{-7}$ MSB, respectively, which are similar but a bit lower than the first method.

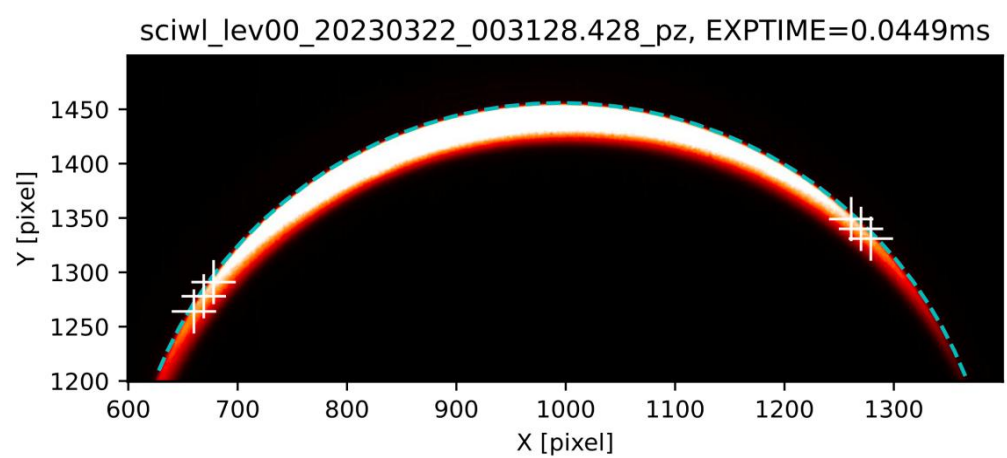

**Figure A1** SCI off-pointing image in which six positions are marked to measure the brightness close to the limb on the solar disk and subsequently calculate the MSB in units of DN.

**B. Estimate of the SCIUV Stray-Light Level**

The imaging wavebands of SCIUV and SDI are both 121.6 nm, and there are overlapping FOV between 1.1 Rs and 1.2 Rs. The brightness of the same prominence is seen by SDI and SCI, so the response relationship between SCIUV and SDI images is established.

Firstly, the same Lyα prominence observed by SCIUV channel and SDI channel at 22:45 UT on 27 February 2023 have been compared each other (see Figure B1). The grayscale level of the prominence from SCIUV is 985.6DN/s after subtracting the neighboring background. The grayscale of prominence from SDI is about 6.96 DN/s after subtracting the neighboring background. Therefore, the grayscale of SCIUV low-gain image is about 141.6 times that of SDI high-gain image after normalizing the exposure time.

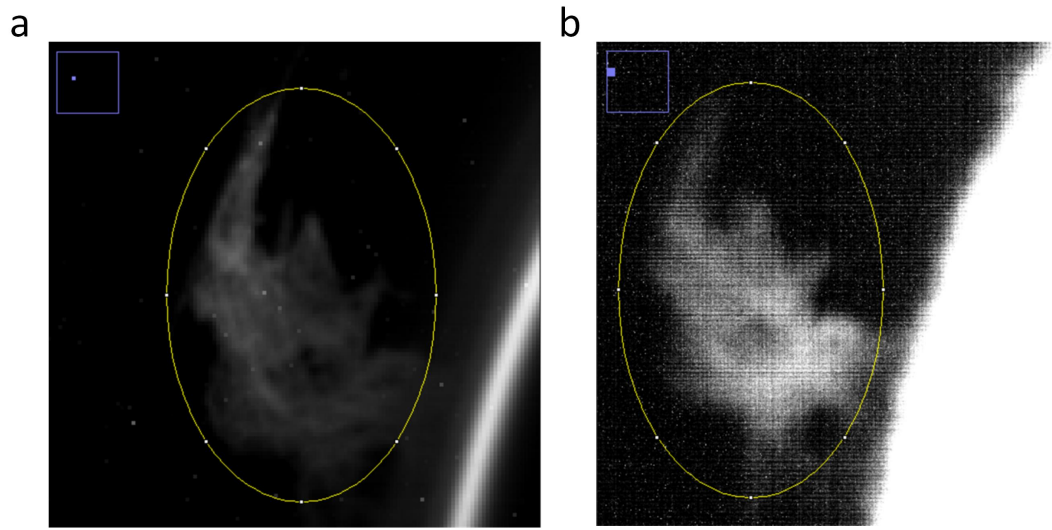

**Figure B1** Prominence observed by SCIUV channel (**a**) and the same prominence observed by SDI channel (**b**).

In Figure B2 a, the grayscale of the coronal image during the solar quiet period of SCIUV on 8 July 2023, is selected to evaluate the stray-light. The grayscale of the image after background subtraction is 20.4 DN/s and 1.98 DN/s respectively at 1.1 Rs and 2.5 Rs. In Figure B2 b, the average grayscale of the normalized time of SDI solar disk image after background removal is 33.8 DN/s  In strict terms, the stray-light level may be evaluated by the minimum grayscale of each pixel corresponding to multiple images when the satellite rotates 45 degrees, and the result calculated by the 45° rotation method is better than that in this paper.

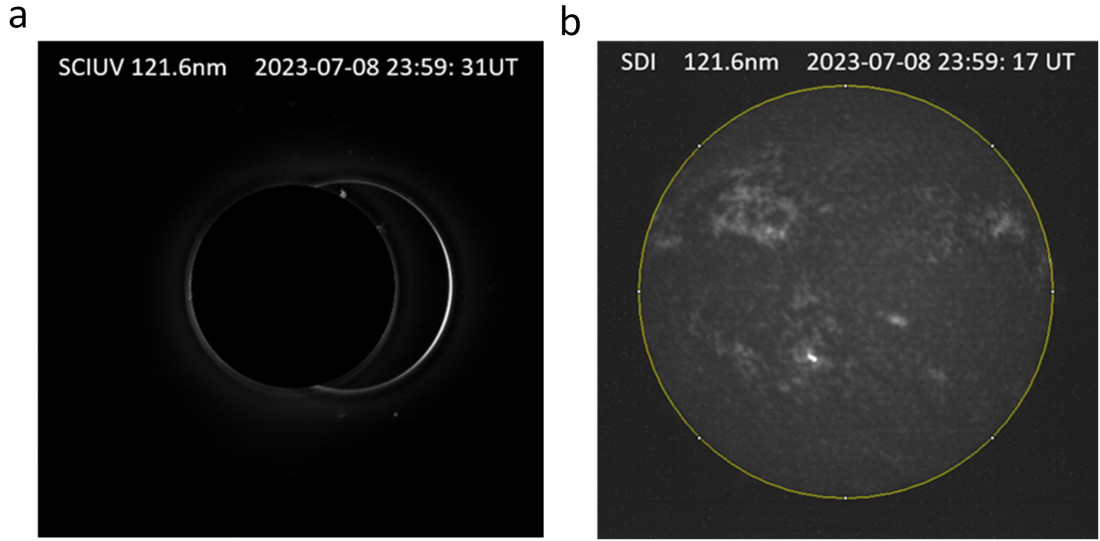

**Figure B2** Images for calculating SCIUV stray-light level (**a**) and the solar disk image by SDI (**b**).

Therefore, the stray-light grayscale of normalized exposure time divided by the average grayscale of the solar disk is the stray-light level. Finally, the stray-light level derived is approximately $4.3 \times 10^{-3}$ MSB and $4.1 \times 10^{-4}$ MSB at 1.1 Rs and 2.5 Rs, respectively.

## C. Estimate of the SDI Spatial Resolution

We evaluate the spatial resolution of the SDI instrument using the fast Fourier transform (FFT) method. We perform two approaches to estimate the resolution, one via the fitting of a Gaussian point spread function (PSF) (Brostrøm and Mølhave, 2022), and the other via the estimate of the signal cutoff based on the amplitude spectrum in the frequency domain. Panel a of Figure C1 presents a full-disk image observed by SDI on 26 October 2022, where a dashed box represents a randomly selected quiet region with a size of 1024×1024 pixels. The zoomed image of this region and its transformed image after FFT in the frequency domain are shown in panel b and c, respectively. We assume the quiet-Sun region in the SDI image can be approximated by the convolution of a step function and a Gaussian PSF. The amplitude spectrum of the SDI image after FFT can be fitted by a convolution function. In panel d, the blue line represents the amplitude of the SDI image in the frequency domain as a function of the normalized frequency (length in units of reciprocal of image size), derived from panel c. The red dashed line represents the optimal fitting curve with a Gaussian full width at half maximum (FWHM) of 18.89±0.29 pixels. For the second method, we calculate the average value $\overline{A}$ of the amplitude spectrum within the frequency range of 0.2 to 0.5, which is used to estimate the noise level ($3 \times \overline{A}$) indicated by the black dashed line. The normalized cutoff frequency of the SDI signal is 0.046, corresponding to a spatial period of 21.79 pixels, which is slightly higher than that derived from the fitting method, as it treats some high-frequency signals as the background noise. Therefore, the spatial resolution we obtained for SDI is around 9.5″.

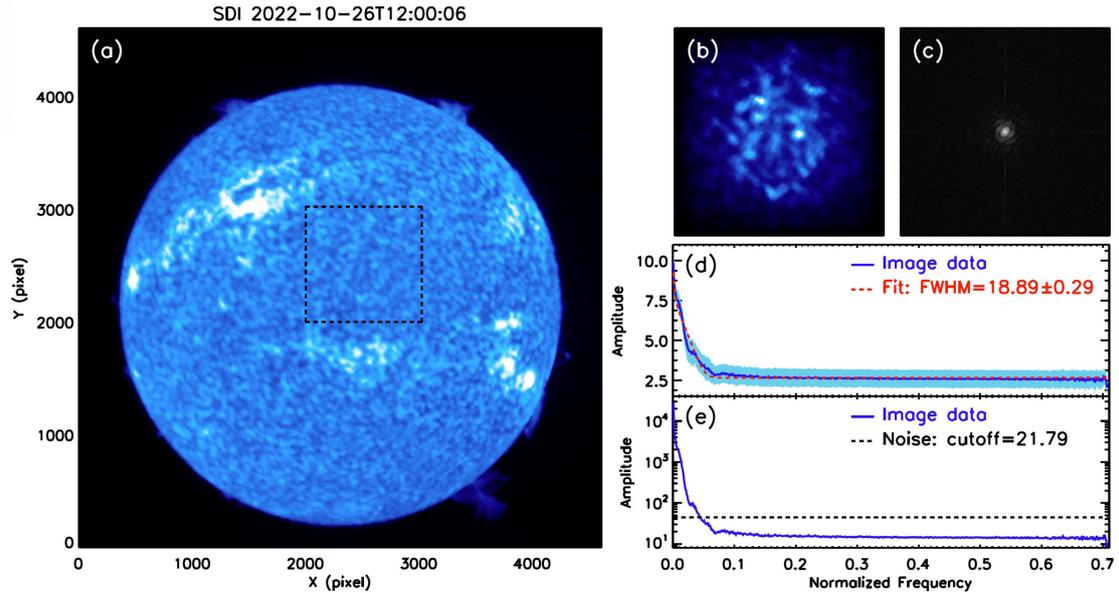

**Figure C1** (**a**) Full-disk image of the SDI observed on 26 October 2022 with a dashed box indicating the selected quiet region with 1024×1024 pixels. (**b**) Zoomed image of SDI filtered using a Hanning window. (**c**) Transformed image in the frequency domain after applying the FFT. (**d**) Results of the fitting of the PSF function: its abscissa and ordinate indicate the normalized frequency (in unit length of the reciprocal of image size) and amplitude, respectively. The blue line represents the amplitude spectrum, and the red dashed line represents the fitting curve. The Gaussian FWHM is about 19 pixels. (e) Results of the SDI signal cutoff. The black line indicates the cutoff level of three times the background noise. The normalized cutoff frequency is around 0.046, corresponding to a spatial period of 21.79 pixels.

## Authors and Affiliations

**Bo Chen[1]· Li Feng[2]· Guang Zhang[1]· Hui Li[2]· Lingping He[1]· Kefei Song[1]· Quanfeng Guo[1]· Ying Li[2]· Yu Huang[2]· Jingwei Li[2]· Jie Zhao[2]· Jianchao Xue[2]· Gen Li[2]· Guanglu Shi[2]· Dechao Song[2]· Lei Lu[2]· Beili Ying· Haifeng Wang[1]· Shuang Dai[1]· Xiaodong Wang[1]· Shilei Mao[1]· Peng**

**Wang**[1]· **Kun Wu**[1]· **Shuai Ren**[1]· **Liang Sun**[1]· **Xianwei Yang**[1]· **Mingyi Xia**[1]· **Xiaoxue Zhang**[1]· **Pen Zhou**[1]· **Chen Tao**[1]· **Yang Liu**[1]· **Sibo Yu**[1]· **Xinka Li**[1]· **Shuting Li**[2]· **Ping Zhang**[2]· **Qiao Li**[2]· **Zhengyuan Tian**[2]· **Yue Zhou**[2]· **Jun Tian**[2]· **Jiahui Shan**[2]· **Xiaofeng Liu**[2]· **Zhichen Jing**[2]· **Weiqun Gan**[2]

✉ B. Chen or L. Feng

chenb@ciomp.ac.cn
lfeng@pmo.ac.cn

[1] Changchun Institute of Optics, Fine Mechanics and Physics, Chinese Academy of Sciences.
[2] Purple Mountain Observatory, Chinese Academy of Sciences.